\documentclass[twocolumn,showpacs,longbibliography,preprintnumbers,amsmath,amssymb]{revtex4-1}
\usepackage{ifpdf}
\usepackage{amsmath}
\usepackage{float}
\usepackage{subfigure}

\ifpdf
    \usepackage[pdftex]{graphicx}
    \usepackage[pdftex,
            plainpages=false,
            breaklinks=true,
            colorlinks=true,
            pdftitle=ice levitation
            pdfauthor=C. Chin
           ]{hyperref}
    \usepackage{thumbpdf}
\else
    \usepackage{graphicx}
    \usepackage{hyperref}
\fi

\begin{document}
\begin{abstract}
We demonstrate levitation and three-dimensionally stable trapping of a wide variety of particles in a vacuum chamber through the use of the thermophoretic force in the presence of a strong temperature gradient. Typical sizes of the trapped particles are between 10~$\mu$m and 1~mm at a pressure between 1 and 10 Torr. The trapping stability is provided by the geometry of the temperature field, as well as the transition between the free molecule and hydrodynamic regimes of the thermophoretic force. To quantitatively measure the thermophoretic force, we examine the levitation heights of spherical polyethylene spheres under various experimental conditions and determine the temperature gradient needed to levitate the particles. A good agreement between our experimental observations and theoretical calculations is obtained. Our system offers a new platform to study thermophoretic phenomena and to simulate dynamics of interacting many-body systems in a microgravity environment.

\end{abstract}
\title{Stable thermophoretic trapping of generic particles at low pressures}
\affiliation{The James Franck Institute, Enrico Fermi Institute and Department of Physics, \\ The University of Chicago, Chicago, Illinois 60637, USA}
\author{Frankie Fung}
\author{Mykhaylo Usatyuk}
\author{B.~J. DeSalvo}
\author{Cheng Chin}
\thanks{Author to whom correspondence should be addressed. Electronic mail: \href{mailto:cchin@uchicago.edu}{cchin@uchicago.edu}}
\date{\today}
\maketitle
\newpage

Levitation of particles in ground-based experiments provides an ideal platform for the study of their dynamics and interactions in a pristine isolated environment, and has broad applications in fundamental and applied sciences. Electromagnetically levitated atoms, ions, and molecules in vacuum have opened the field of quantum degenerate gases \cite{Anderson95, Davis95}, cold chemistry \cite{Krems08}, and atom-based quantum information \cite{Kielpinski02}. Optical levitation of micron-size objects \cite{Ashkin71} provides a new playground to investigate aerosols in atmosphere \cite{Mund03} and Brownian motion of microspheres \cite{Kheifets14}.

Levitation of macroscopic particles in low pressure is of particular interest due to its wide applications in space, atmospheric, and even astro-chemical research. Several schemes to levitate glass spheres and water droplets have been realized based on optical tweezers \cite{Ashkin75}, high magnetic field gradients \cite{Berry97,Ikezoe98}, and electrodynamic balance \cite{Dhariwal93}. Experiments on dusty plasma further show many-body dynamics of levitating dust particles. An interesting scheme was demonstrated recently to levitate generic particles above a hot surface by 100~$\mu$m based on the Knudsen compressor effect \cite{Kelling09,Knudsen09}. The authors further demonstrated levitation of ice particles 1~cm above a hot surface for a few seconds \cite{Kelling11}.

\begin{figure}[htbp]
\centering
\includegraphics[width=.42\textwidth]{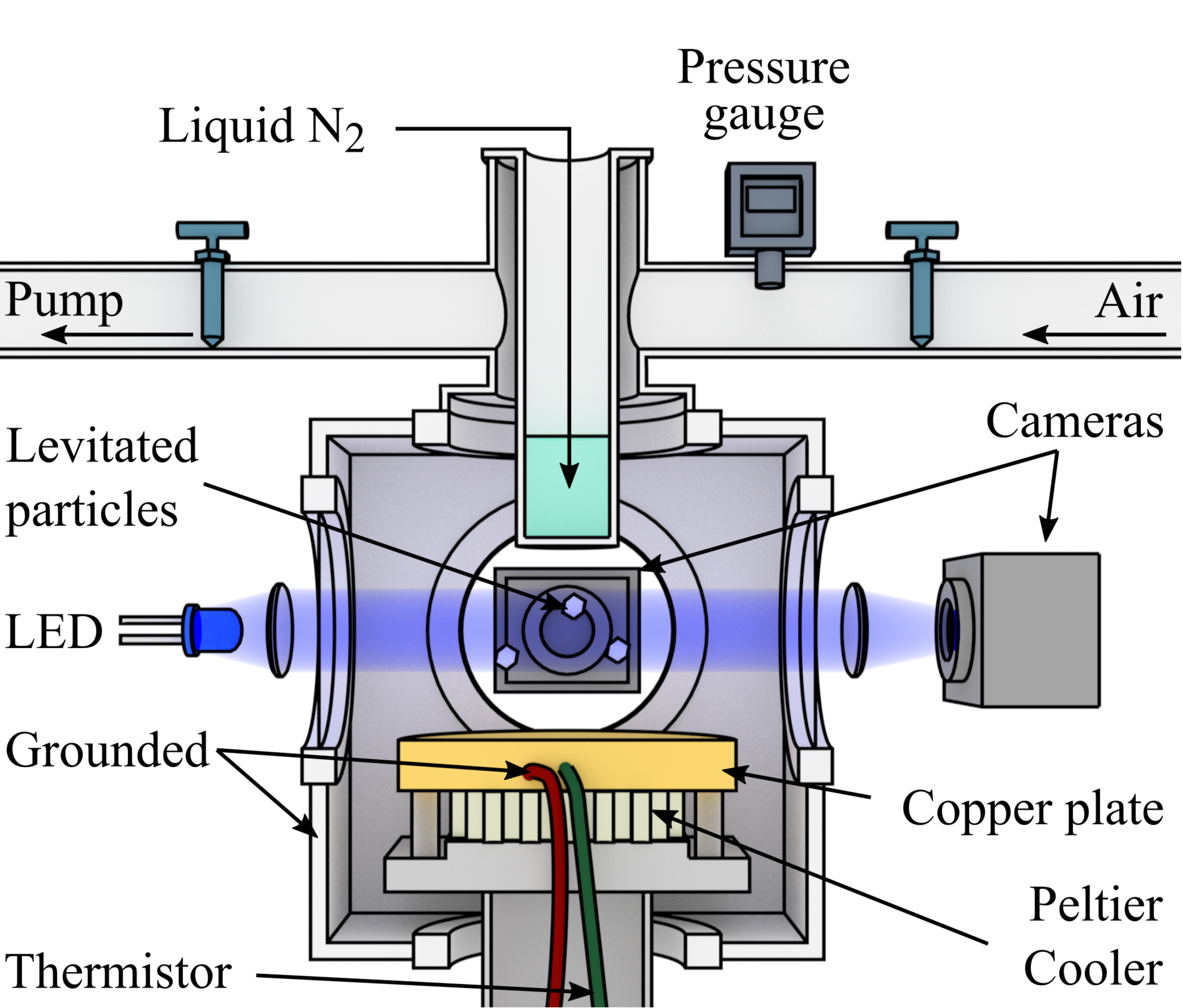}
\caption{Experimental apparatus. The copper plate and liquid nitrogen bucket establish a temperature gradient. The pressure is regulated by a vacuum pump and a needle valve that opens to atmosphere. Temperature regulation of the copper plate is achieved with a Peltier cooler. An absorption imaging setup using collimated blue LED light captures the shape of levitating particles. A second camera records their dynamics. During the levitation, we continuously monitor the temperatures of the top and bottom plates. The plates and the chamber are electrically grounded to avoid electrostatic forces on the levitating particles.}
\label{fig:1}
\end{figure}

In this paper, we report an experimental setup to achieve stable, three-dimensional trapping of various particles at pressures of $\mathit{P}$=1$\sim$10~Torr. The levitation and trapping do not require external electromagnetic forces or the Knudsen compressor effect, and are achieved with only the thermophoretic force in the presence of a strong temperature gradient. We observe stable trapping of various objects including ice particles, ceramic spheres, hollow glass bubbles, polyethylene spheres, and lint. The stability of levitation allows us to measure the levitation heights of the particles as well as the dynamics. Based on a numerical model to describe the temperature distribution in our system, we obtain a good agreement between our experimental observation and theoretical calculation over a large range of temperatures where stable levitation is achieved.

Our experimental setup is displayed in Fig.~\ref{fig:1}. At the center of the vacuum chamber is a 10-mm gap between a warm copper plate and a stainless steel bucket filled with liquid nitrogen at 77~K. The bottom copper plate has a diameter of 33.7~mm and is attached to a Peltier cooler for temperature control. A thermistor is embedded in the plate to allow for temperature monitoring. The stainless steel bucket has a smaller outer diameter of 25.4~mm. The bottom surface of the bucket acts as the top plate and its temperature is monitored by a K-type thermocouple. In the presence of a large temperature gradient, particles are levitated and trapped near the mid-plane between the the top and bottom plates.

\begin{figure}[htbp]
\centering
\includegraphics[width=.48\textwidth]{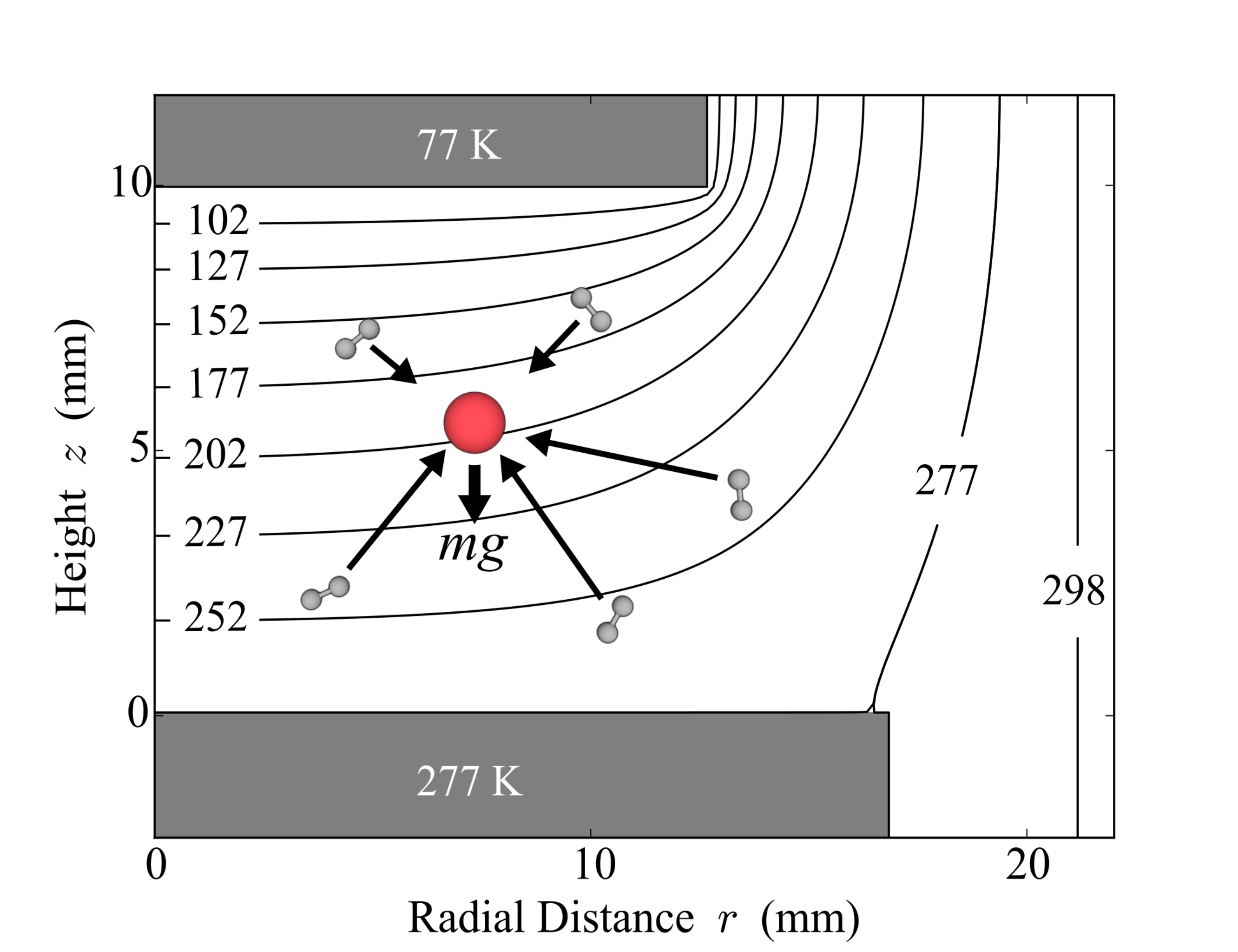}
\caption{Simulation of temperature distribution in the chamber is obtained by solving Eq.~(\ref{eq:2}). Boundary conditions hold the top plate at 77~K, bottom plate at 277~K, and chamber wall at 298~K. Temperature contours are labeled. In the presence of a temperature gradient, collisions with background gas particles exert a net force opposite the direction of the temperature gradient owing to the larger momentum transfer from hotter molecules.  Due to our geometry, particles experience both an upward force counteracting gravity $\mathit{mg}$ and a radial confining force.}
\label{fig:2}
\end{figure}

The levitation comes from the thermophoretic force in our chamber. As illustrated in Fig.~\ref{fig:2}, air molecules coming from the hotter side transfer on average a larger momentum to the particle than those from the colder side. A net momentum transfer due to collisions with molecules overcomes gravity and pulls the particles toward the symmetry axis of the two plates.

In the presence of a temperature gradient $\nabla T$, the thermophoretic force acting on a sphere of radius $a$ is given by \cite{Zheng02},

\begin{eqnarray}
F_{th}&=&- f_T \frac{a^2 \kappa}{v} \nabla T,
\label{eq:1}
\end{eqnarray}

\noindent where $\kappa$ is the thermal conductivity of the medium (rarefied air in our case), $v=\sqrt{2k_{\text{B}}T/m}$ is the most probable speed of the air molecules, $m$ is the molecular mass, $k_{\text{B}}$ is the Boltzmann constant, and $f_T$ is a dimensionless thermophoretic parameter that depends on the Knudsen number Kn$=\ell/a$. The Knudsen number is defined as the ratio between the molecular mean free path $\ell=\mu v/P$ and $a$, where $\mu$ is the dynamic viscosity. In the free molecule regime where Kn$\gg 1$, $f_T$ reaches the maximum value of $16\sqrt{\pi}/15\approx1.89$. In the hydrodynamic limit where Kn$\ll1$, however, $f_T$ approaches zero and the thermophoretic force vanishes \cite{Zheng02}.

The thermophoretic force described in Eq.~(\ref{eq:1}) is associated with heat transfer in the thermal conduction regime. In a temperature field $T(\vec{x})$, heat flows according to Fourier's law as $Q=-\kappa \nabla T$. Associated with the heat flow, momentum flow per unit area can be estimated as $\dot{p}/A = Q/v$, where $A$ is the effective cross section of a particle. The force on a sphere with radius $a$ is thus $F = \dot{p} = -\pi (a^2 \kappa/v) \nabla T$, and is consistent with Eq.~(\ref{eq:1}) in the molecular flow regime.

To stably levitate and trap a particle, one requires a restoring force toward the trap center for small displacements in all directions. It may appear that a three-dimensional minimum in the temperature field $T(\vec{x})$ is required. Such condition, however, cannot be satisfied without a heat sink under the condition that $\nabla\cdot Q=0$ (Earnshaw's theorem). Together with Fourier's law, the temperature field in free space satisfies

\begin{eqnarray}
\nabla\cdot(\kappa \nabla T)=0.
\label{eq:2}
\end{eqnarray}

\noindent Since the thermal conductivity of air increases with temperature \cite{Kadoya85}, the above equation gives  $\nabla^2 T=- |\nabla T|^2 \kappa'(T)/\kappa(T)< 0$, which excludes the possibility of stably confining a particle in free space based solely on a temperature gradient.

The requirement for stable thermophoretic levitation is given by $\nabla\cdot (F_{th}+F_G)<0$, where the $F_G$ is the gravitational force. Using Eqs.~(\ref{eq:1}), (\ref{eq:2}), and $\nabla\cdot F_G=0$ in free space, we obtain the necessary condition for stable levitation:

\begin{eqnarray}
-\nabla \cdot F_{th}=\kappa a^2 |\nabla T|^2\frac{d}{dT}\frac{f_T}{v}>0.
\label{eq:3}
\end{eqnarray}

\noindent Since thermal conductivity $\kappa$ is positive, the above condition is equivalent to $d(f_T/\sqrt{T})/dT>0$.

\begin{figure}[htbp]
\centering
\includegraphics[width=.48\textwidth]{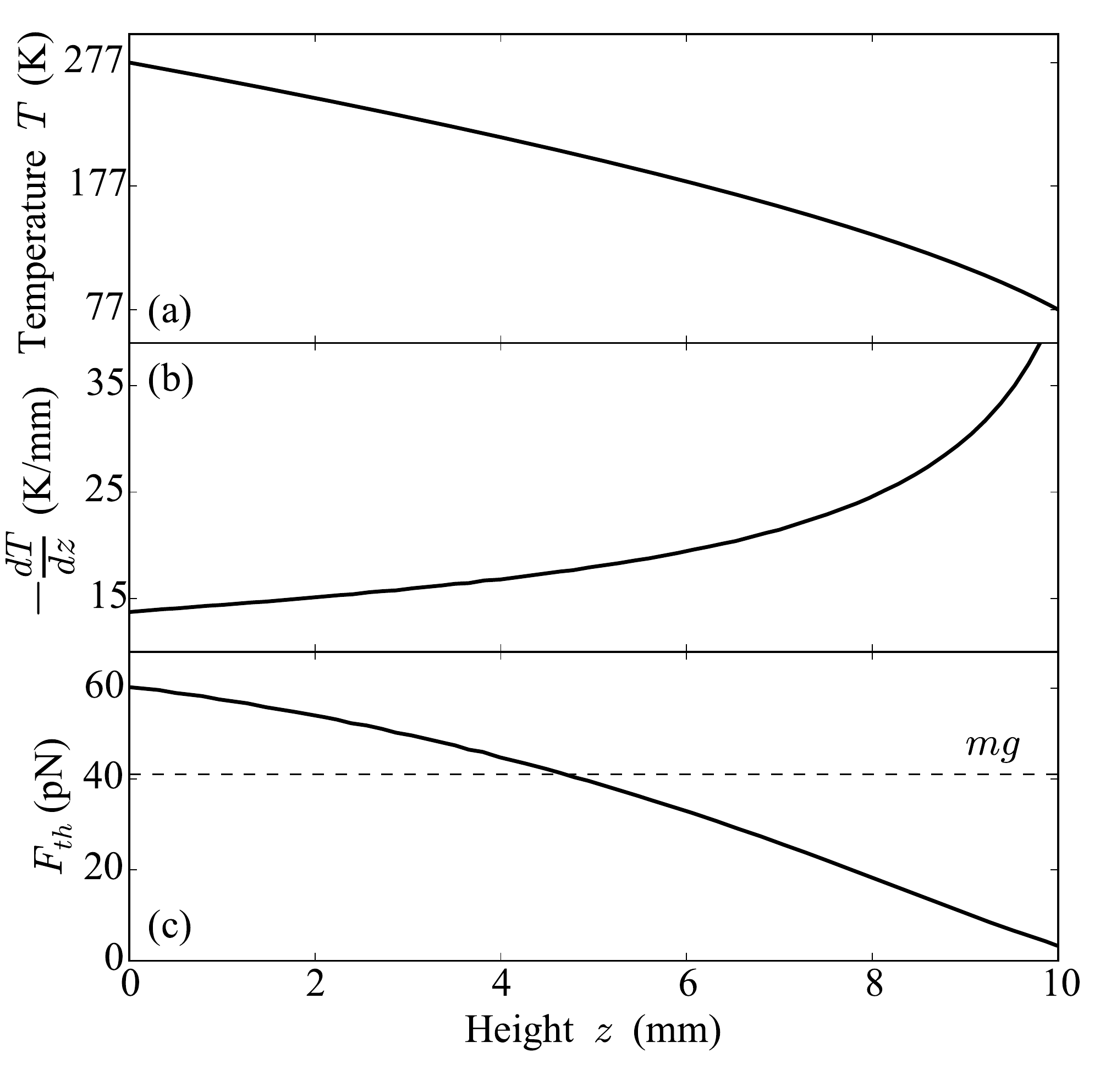}
\caption{Thermodynamic quantities and thermophoretic force simulation for a 10 $\mu$m polyethylene sphere with $\rho$=998 kg/$m^{3}$ and $\mathit{P}$=6 Torr. Using results in Fig.~\ref{fig:2}, the temperature shown in (a) and temperature gradient shown in (b) are plotted as functions of height. (c) Thermophoretic force (solid line), calculated from Eq.~(\ref{eq:1}), is compared with the gravitational force on the particle (dashed line). The intersection of the two curves indicates the position of the levitating particle. The weaker levitation force as the particle rises offers the levitation stability.}
\label{fig:3}
\end{figure}

An example of the temperature distribution in our system is shown in Fig.~\ref{fig:2} and is numerically calculated based on Eq.~(\ref{eq:2}). Here one can see a positive curvature of temperature in the radial direction that offers radial confinement. In the vertical direction, however, the curvature of the temperature is negative, shown in more clearly in Fig.~\ref{fig:3}(a).

The key to stabilizing particles in the vertical direction is the thermophoretic parameter $f_T$, which is strongly suppressed when particles enter the hydrodynamic regime. Since our system is isobaric $p=nk_{B}T=const.$, lower temperature near the colder top plate suggests higher density. Particles moving near the top plate are thus deeper in the hydrodynamic regime with a corresponding smaller thermophoretic force. The total force thus switches sign when the particles are over-levitated. The distribution of temperature, temperature gradient, and levitation force in our system for a typical operation are shown in Fig.~\ref{fig:3}.

\begin{figure}[htbp]
\centering
\includegraphics[width=.50\textwidth]{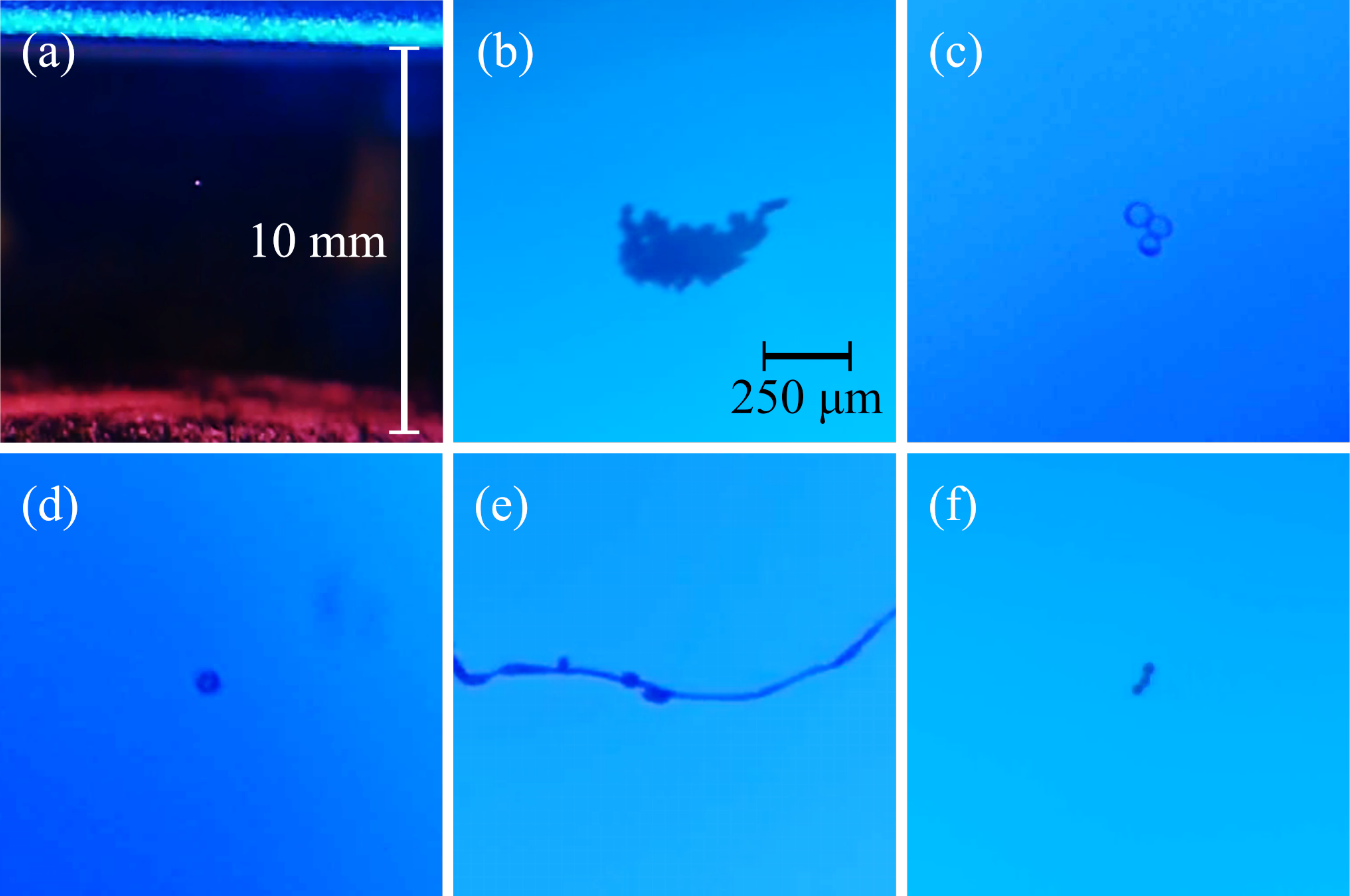}
\caption{Gallery of levitating particles. (a) Low magnification image of one polyethylene sphere (Cospheric, UVPMS-BR-0.995). High magnification images of (b) one ice particle, (c) an aggregate of three hollow glass bubbles (Miapoxy, 406354), (d) one ceramic sphere (Miapoxy, 406344), (e) one piece of lint, and (f)  an aggregate of three polyethylene spheres.}
\label{fig:4}
\end{figure}

The experimental procedure goes as follows. We first deposit the desired particles on the bottom copper plate. After sealing the chamber, we pump the chamber down to $\mathit{P}$ = 1$\sim$10 Torr. A pressure gauge (Thyracont, VD8) continuously records the pressure throughout the experiment. We then add liquid nitrogen to the steel bucket and the top plate quickly reaches 77~K. We launch the particles by vibrating the chamber, and once above the bottom plate the particles experience the upward thermophoretic force. Depending on the strength of vibration, the pressure, and the type of particles, the number of levitated particles ranges from one to hundreds. Typically ceramic spheres and hollow glass bubbles tend to be levitated in greater numbers than polyethylene spheres. Lint is occasionally introduced into the chamber, and we observe stably levitating lint as long as 1~mm in length. Fig~\ref{fig:4} displays a gallery of levitating objects of different types.

The levitation procedure for ice particles differs from the above. The top plate is cooled down to 77~K before we seal the chamber. Water vapor condensing on the top plate creates a layer of ice particles. The chamber is then pumped down to the desired pressure. Upon external vibration of the chamber ice particles fall from the top plate and some are levitated.

Our detection system consists of two digital cameras along with a light-emitting diode (LED) that illuminates the levitating particles. One camera with high magnification faces the collimated LED beam and provides a closeup view of the particles. Representative images are shown in Fig.~\ref{fig:4} (b)-(f). The second camera records the dynamics of particles with a full field of view including the top and bottom plates, shown in Fig.~\ref{fig:4} (a).

To quantitatively characterize the levitating force at play, we analyze the measured height of singly levitating polyethylene spheres. We focus on single particle levitation to avoid the complex dynamics of multiple particles. In this case, particles remain stably levitated for our longest observation times of one hour, giving us ample time to study the conditions of levitation. To aid comparison with theoretical calculations \cite{Zheng02}, we focus on spherical particles. After examination with a microscope, polyethylene spheres were chosen as they appeared the closest to ideal spheres.

Since levitating particles are easily affected by air currents resulting from pressure changes, we keep pressure constant and vary the temperature of the top plate. To achieve this we allow the liquid nitrogen in the steel bucket to evaporate, which slowly warms up the top plate over 5$\sim$10 minutes. During this time, we synchronously record the temperature of the plates, pressure, and particle height.

\begin{figure}[htbp]
\includegraphics[width=.5\textwidth]{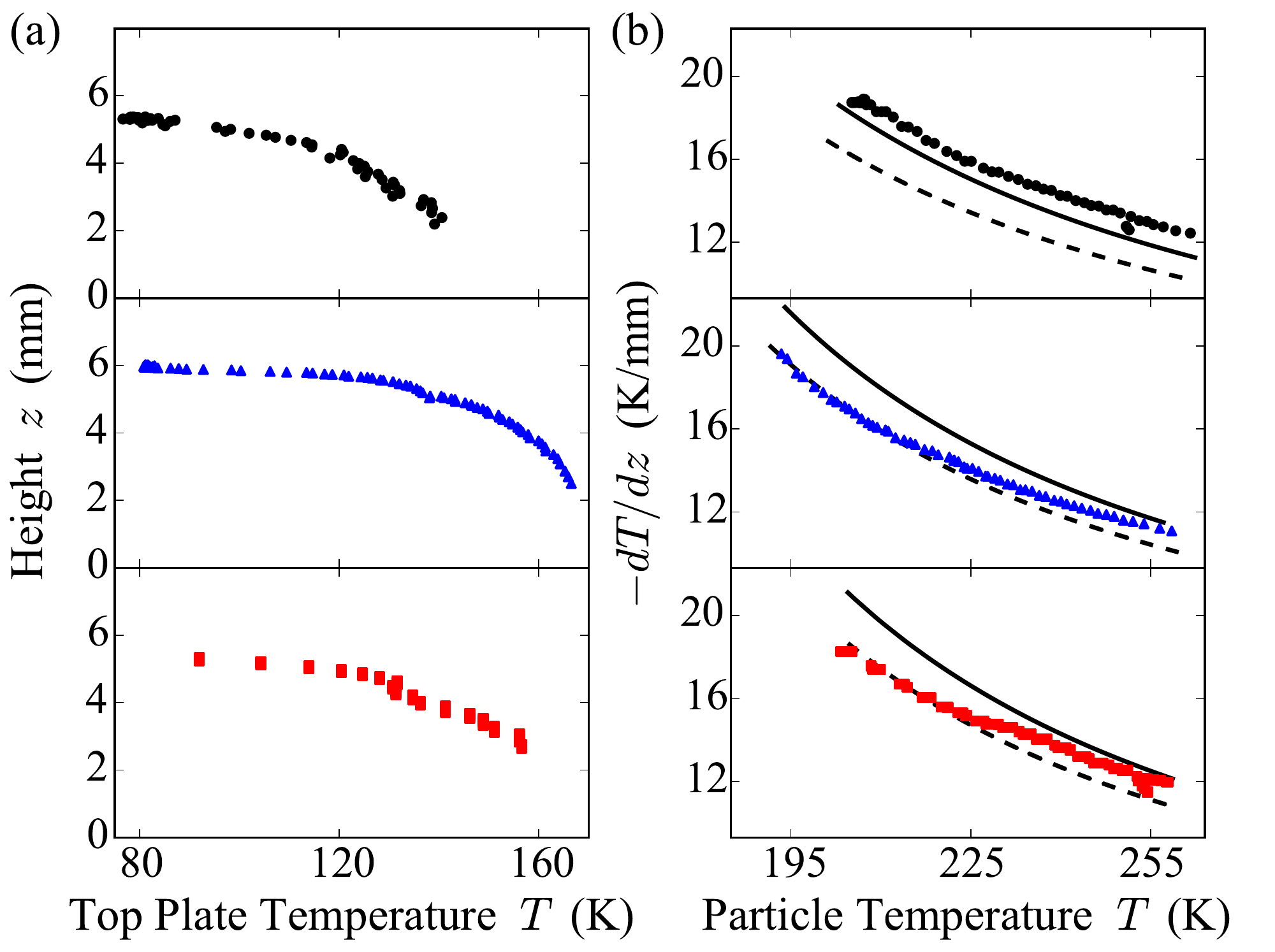}
\caption[Caption for LOF]{Dynamics of levitating polyethylene spheres during a temperature increase. (a) Levitation heights of single polyethylene spheres. After levitation we allow the top plate to warm up and record the levitation height. Circles (black) correspond to a sphere of radius $\mathit{a}$ = 12.7 $\mu$m at $\mathit{P}$ = 3.9 Torr, triangles (blue) correspond to a particle of $\mathit{a}$ =9.8 $\mu$m at $\mathit{P}$ = 6.4 Torr, and squares (red) correspond to a particle of radius 9.3 $\mu$m at $\mathit{P}$ = 7.3 Torr. (b) Temperatures and temperature gradients at the location of particles obtained by simulation of temperature distribution in the chamber. Theoretical predictions from Takata (solid line)\cite{Takata94} and Yamamoto (dashed line)\cite{Yamamoto88} are plotted for comparison.}
\label{fig:5}
\end{figure}

Examples of singly levitating polyethylene spheres are shown in Fig.~\ref{fig:5}, where the top plate temperature slowly increases from 77~K toward room temperature. The particle's levitation height gradually decreases as the top plate warms up because of the decreasing temperature gradient in the chamber.

Stable levitation is characterized by the balance of gravity and the thermophoretic force:

\begin{eqnarray}
F_{th}&=&\rho Vg,
\label{eq:4}
\end{eqnarray}

\noindent where $\rho=998~$kg/m$^3$ is the density of polyethylene spheres, g = 9.8~m/s is gravitational acceleration, and $V$ is particle volume determined from the particle radius. To calculate the thermophoretic force we determine the thermodynamic quantities at the location of the particle, and extract the thermophoretic parameter $f_{T}$ for a specific Knudsen number Kn from Eqs.~(\ref{eq:1}) and (\ref{eq:3}).

\begin{figure}[htbp]
\centering
\includegraphics[width=.48\textwidth]{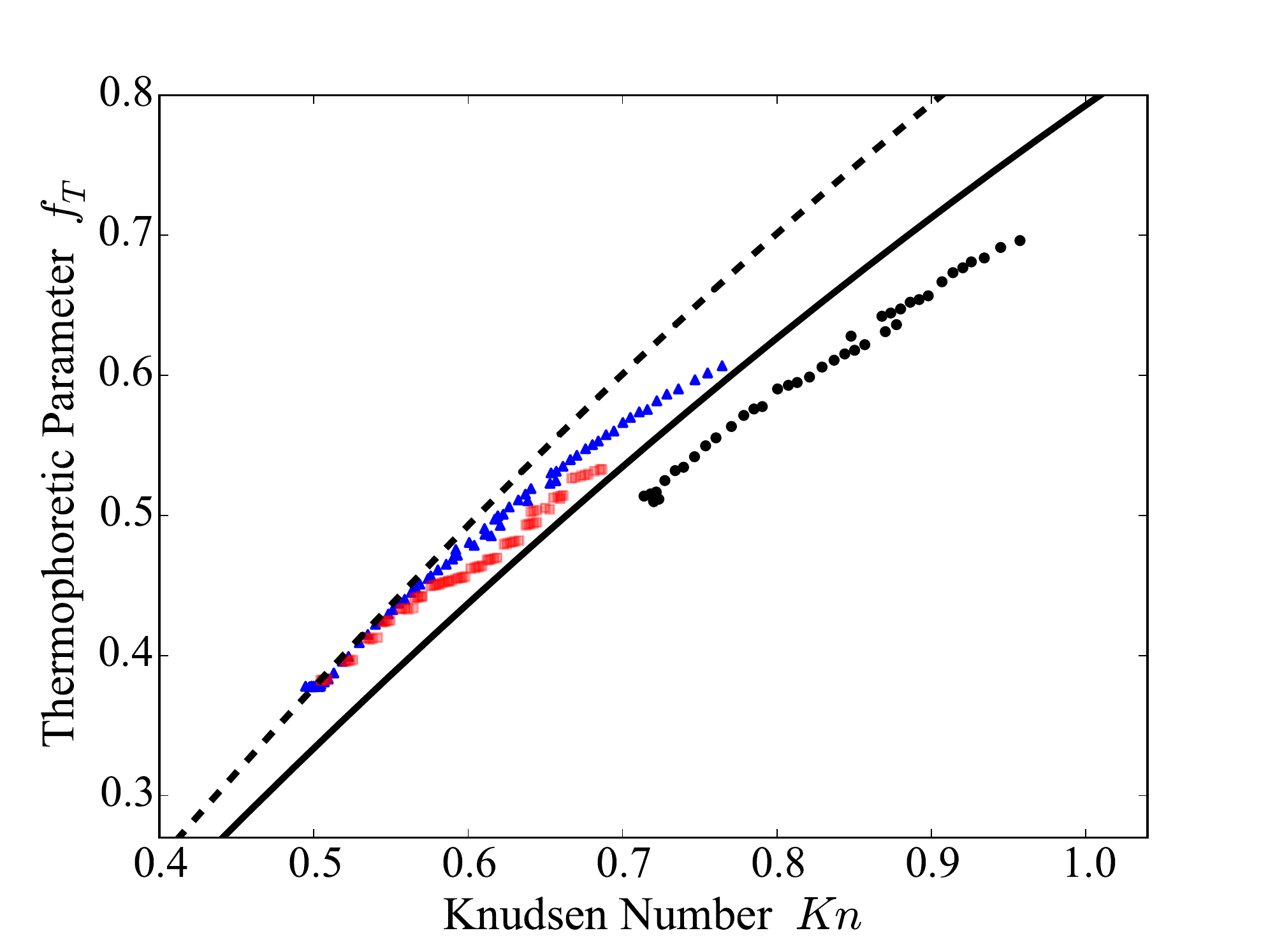}
\caption{Comparison between $f_{T}$ from experimental measurements and theoretical models. Thermophoretic parameter $f_{T}$ is calculated from the data shown in Fig.~\ref{fig:5}. The solid line represents interpolation of the theoretical calculation of $f_{T}$ by Takata et al. \cite{Takata94}. The dashed line represents interpolation of the theoretical calculation of $f_{T}$ by Yamamoto et al. \cite{Yamamoto88} with $k_{12}=0.03$, where $k_{12}$ is the ratio of the thermal conductivity of gas and the levitating particle.}
\label{fig:6}
\end{figure}

To determine the thermodynamic quantities of air at the particle's location, we solve the air temperature distribution in the chamber using the steady state heat diffusion equation, Eq.~(\ref{eq:2}), with the appropriate boundary conditions. From the distribution we compute the temperature and temperature gradient at the location of the particle. To obtain the thermal conductivity $\kappa$ and dynamic viscosity $\mu$ of rarefied air, we obtained their functional forms by fitting the data in Ref.~\cite{Kadoya85} and evaluate them at the position of the particle. We also calculate the Knudsen number Kn$=\ell/a$ at the particle position. Finally, we determine the value of $f_{T}$ for Eq.~(\ref{eq:3}) to hold. The procedure is repeated during the entire warming up process. The measurements on three independent polyethylene spheres are shown in Fig.~\ref{fig:6}.

Our measurements are consistent with the theories of Takata et al. \cite{Takata94} and Yamamoto et al. \cite{Yamamoto88} within experimental uncertainties. The discrepancies of $5\sim10\%$ between the measurements on three particles are possibly caused by systematic uncertainties of our experiment. The particle radii determination suffers from an uncertainty of approximately 1 $\mu$m due to our imaging resolution. Furthermore, the two plates do not have a uniform temperature distribution, and the particles can levitate slightly away from the symmetry axis. These result in uncertainties of temperature and temperature gradient at the location of the particle less than 10$\%$.

Another systematic effect is the photophoretic force coming from external radiation. Such force is significant in former levitation experiments ~\cite{Kelling11, Eckerskorn15}. We estimate the photophoretic force in our system based on Refs.~\cite{Rohatschek95, Eckerskorn15} and find that the radiation power of 10~W/m$^2$ from the copper plate at 277~K and 2~W/m$^2$ from the stainless steel bucket at 77~K contribute a net levitation force of $<0.03$~pN on the particles. Compared to the typical particle weights of 25$\sim$100~pN, the photophoretic force is negligible.

Convective flow is another potential systematic. Convective flow occurs due to the Rayleigh-B\'{e}nard instability, which occurs in a rarified gas at high Rayleigh number given by \cite{Golshtein1996,Stefanov2002}

\begin{eqnarray}
\mbox{Ra}=\frac{2048}{75\pi}\frac{1-r}{(1+r)^2}\frac1{\mbox{Fr}{\mbox{Kn}^*}^2}>1708.
\label{eq:5}
\end{eqnarray}

\noindent Here $\mbox{Kn}^*=\ell/H$ is the Knudsen number of the system, Fr$=v^2/gH$ is the Froude number, $r=T_c/T_h$ is the temperature ratio of the two plates, and $H$ is plate separation. Given typical parameters of our system $T_c=77~K$, $T_h=277~K$, $\ell=7~\mu$m, $H=1$~cm, and $v=290$~m/s, our Rayleigh number Ra$=9.1$ is well below the critical value of 1708. We thus conclude that convection is insignificant.

In conclusion we demonstrate a versatile experimental system that levitates particles of various types, and stably confines them. We perform a quantitative study on singly levitating polyethylene spheres and establish a numerical model to determine the thermodynamic quantities at the location of the particles. A direct comparison between experiment and theory on the levitation condition allows us to extract the thermophoretic parameter in the range of Knudsen number $0.5<$Kn$<1.0$. We report good agreement with theoretical models.

Our system offers a new platform to study levitation and interactions of generic particles. When two and more particles are levitated, we observe interesting dynamics in the presence of interparticle interactions and external forces. Our system also offers the possibility to simulate dynamics of macroscopic particles in a microgravity environment, enhancing our capability to detect feeble forces between macroscopic objects in ground-based laboratories.

We thank N. Kowalski, Bernard Xie, and C. Parker for assistance in the early phase of the experiment and acknowledge useful discussions with H. Jaeger, S. Nagel and D. Schuster. This work was supported by NSF MRSEC Grant No. DMR-1420709.

\bibliography{BibDatabase}

\begin{thebibliography}{22}%
\makeatletter
\providecommand \@ifxundefined [1]{%
 \@ifx{#1\undefined}
}%
\providecommand \@ifnum [1]{%
 \ifnum #1\expandafter \@firstoftwo
 \else \expandafter \@secondoftwo
 \fi
}%
\providecommand \@ifx [1]{%
 \ifx #1\expandafter \@firstoftwo
 \else \expandafter \@secondoftwo
 \fi
}%
\providecommand \natexlab [1]{#1}%
\providecommand \enquote  [1]{``#1''}%
\providecommand \bibnamefont  [1]{#1}%
\providecommand \bibfnamefont [1]{#1}%
\providecommand \citenamefont [1]{#1}%
\providecommand \href@noop [0]{\@secondoftwo}%
\providecommand \href [0]{\begingroup \@sanitize@url \@href}%
\providecommand \@href[1]{\@@startlink{#1}\@@href}%
\providecommand \@@href[1]{\endgroup#1\@@endlink}%
\providecommand \@sanitize@url [0]{\catcode `\\12\catcode `\$12\catcode
  `\&12\catcode `\#12\catcode `\^12\catcode `\_12\catcode `\%12\relax}%
\providecommand \@@startlink[1]{}%
\providecommand \@@endlink[0]{}%
\providecommand \url  [0]{\begingroup\@sanitize@url \@url }%
\providecommand \@url [1]{\endgroup\@href {#1}{\urlprefix }}%
\providecommand \urlprefix  [0]{URL }%
\providecommand \Eprint [0]{\href }%
\providecommand \doibase [0]{http://dx.doi.org/}%
\providecommand \selectlanguage [0]{\@gobble}%
\providecommand \bibinfo  [0]{\@secondoftwo}%
\providecommand \bibfield  [0]{\@secondoftwo}%
\providecommand \translation [1]{[#1]}%
\providecommand \BibitemOpen [0]{}%
\providecommand \bibitemStop [0]{}%
\providecommand \bibitemNoStop [0]{.\EOS\space}%
\providecommand \EOS [0]{\spacefactor3000\relax}%
\providecommand \BibitemShut  [1]{\csname bibitem#1\endcsname}%
\let\auto@bib@innerbib\@empty
\bibitem [{\citenamefont {Anderson}\ \emph {et~al.}(1995)\citenamefont
  {Anderson}, \citenamefont {Ensher}, \citenamefont {Matthews}, \citenamefont
  {Wieman},\ and\ \citenamefont {Cornell}}]{Anderson95}%
  \BibitemOpen
  \bibfield  {author} {\bibinfo {author} {\bibfnamefont {M.~H.}\ \bibnamefont
  {Anderson}}, \bibinfo {author} {\bibfnamefont {J.~R.}\ \bibnamefont
  {Ensher}}, \bibinfo {author} {\bibfnamefont {M.~R.}\ \bibnamefont
  {Matthews}}, \bibinfo {author} {\bibfnamefont {C.~E.}\ \bibnamefont
  {Wieman}}, \ and\ \bibinfo {author} {\bibfnamefont {E.~A.}\ \bibnamefont
  {Cornell}},\ }\bibfield  {title} {\enquote {\bibinfo {title} {Observation of
  {B}ose-{E}instein condensation in a dilute atomic vapor},}\ }\href@noop {}
  {\bibfield  {journal} {\bibinfo  {journal} {Science}\ }\textbf {\bibinfo
  {volume} {269}},\ \bibinfo {pages} {198--201} (\bibinfo {year}
  {1995})}\BibitemShut {NoStop}%
\bibitem [{\citenamefont {Davis}\ \emph {et~al.}(1995)\citenamefont {Davis},
  \citenamefont {Mewes}, \citenamefont {Andrews}, \citenamefont {van Druten},
  \citenamefont {Durfee}, \citenamefont {Kurn},\ and\ \citenamefont
  {Ketterle}}]{Davis95}%
  \BibitemOpen
  \bibfield  {author} {\bibinfo {author} {\bibfnamefont {K.~B.}\ \bibnamefont
  {Davis}}, \bibinfo {author} {\bibfnamefont {M.~O.}\ \bibnamefont {Mewes}},
  \bibinfo {author} {\bibfnamefont {M.~R.}\ \bibnamefont {Andrews}}, \bibinfo
  {author} {\bibfnamefont {N.~J.}\ \bibnamefont {van Druten}}, \bibinfo
  {author} {\bibfnamefont {D.~S.}\ \bibnamefont {Durfee}}, \bibinfo {author}
  {\bibfnamefont {D.~M.}\ \bibnamefont {Kurn}}, \ and\ \bibinfo {author}
  {\bibfnamefont {W.}~\bibnamefont {Ketterle}},\ }\bibfield  {title} {\enquote
  {\bibinfo {title} {Bose-{E}instein condensation in a gas of sodium atoms},}\
  }\href@noop {} {\bibfield  {journal} {\bibinfo  {journal} {Phys. Rev. Lett.}\
  }\textbf {\bibinfo {volume} {75}},\ \bibinfo {pages} {3969--3973} (\bibinfo
  {year} {1995})}\BibitemShut {NoStop}%
\bibitem [{\citenamefont {Krems}(2008)}]{Krems08}%
  \BibitemOpen
  \bibfield  {author} {\bibinfo {author} {\bibfnamefont {R.~V.}\ \bibnamefont
  {Krems}},\ }\bibfield  {title} {\enquote {\bibinfo {title} {Cold controlled
  chemistry},}\ }\href@noop {} {\bibfield  {journal} {\bibinfo  {journal}
  {Phys. Chem. Chem. Phys.}\ }\textbf {\bibinfo {volume} {10}},\ \bibinfo
  {pages} {4079--4092} (\bibinfo {year} {2008})}\BibitemShut {NoStop}%
\bibitem [{\citenamefont {Kielpinski}\ \emph {et~al.}(2002)\citenamefont
  {Kielpinski}, \citenamefont {Monroe},\ and\ \citenamefont
  {Wineland}}]{Kielpinski02}%
  \BibitemOpen
  \bibfield  {author} {\bibinfo {author} {\bibfnamefont {D.}~\bibnamefont
  {Kielpinski}}, \bibinfo {author} {\bibfnamefont {C.}~\bibnamefont {Monroe}},
  \ and\ \bibinfo {author} {\bibfnamefont {D.~J.}\ \bibnamefont {Wineland}},\
  }\bibfield  {title} {\enquote {\bibinfo {title} {Architecture for a
  large-scale ion-trap quantum computer},}\ }\href@noop {} {\bibfield
  {journal} {\bibinfo  {journal} {Nature}\ }\textbf {\bibinfo {volume} {417}},\
  \bibinfo {pages} {709--711} (\bibinfo {year} {2002})}\BibitemShut {NoStop}%
\bibitem [{\citenamefont {Ashkin}\ and\ \citenamefont
  {Dziedzic}(1971)}]{Ashkin71}%
  \BibitemOpen
  \bibfield  {author} {\bibinfo {author} {\bibfnamefont {A.}~\bibnamefont
  {Ashkin}}\ and\ \bibinfo {author} {\bibfnamefont {J.~M.}\ \bibnamefont
  {Dziedzic}},\ }\bibfield  {title} {\enquote {\bibinfo {title} {Optical
  levitation by radiation pressure},}\ }\href@noop {} {\bibfield  {journal}
  {\bibinfo  {journal} {Appl. Phys. Let.}\ }\textbf {\bibinfo {volume} {19}},\
  \bibinfo {pages} {283--285} (\bibinfo {year} {1971})}\BibitemShut {NoStop}%
\bibitem [{\citenamefont {Mund}\ and\ \citenamefont {Zellner}(2003)}]{Mund03}%
  \BibitemOpen
  \bibfield  {author} {\bibinfo {author} {\bibfnamefont {C.}~\bibnamefont
  {Mund}}\ and\ \bibinfo {author} {\bibfnamefont {R.}~\bibnamefont {Zellner}},\
  }\bibfield  {title} {\enquote {\bibinfo {title} {Optical levitation of single
  microdroplets at temperatures down to 180~{K}},}\ }\href@noop {} {\bibfield
  {journal} {\bibinfo  {journal} {ChemPhysChem}\ }\textbf {\bibinfo {volume}
  {4}},\ \bibinfo {pages} {630--638} (\bibinfo {year} {2003})}\BibitemShut
  {NoStop}%
\bibitem [{\citenamefont {Kheifets}\ \emph {et~al.}(2014)\citenamefont
  {Kheifets}, \citenamefont {Simha}, \citenamefont {Melin}, \citenamefont
  {Li},\ and\ \citenamefont {Raizen}}]{Kheifets14}%
  \BibitemOpen
  \bibfield  {author} {\bibinfo {author} {\bibfnamefont {S.}~\bibnamefont
  {Kheifets}}, \bibinfo {author} {\bibfnamefont {A.}~\bibnamefont {Simha}},
  \bibinfo {author} {\bibfnamefont {K.}~\bibnamefont {Melin}}, \bibinfo
  {author} {\bibfnamefont {T.}~\bibnamefont {Li}}, \ and\ \bibinfo {author}
  {\bibfnamefont {M.~G.}\ \bibnamefont {Raizen}},\ }\bibfield  {title}
  {\enquote {\bibinfo {title} {Observation of brownian motion in liquids at
  short times: Instantaneous velocity and memory loss},}\ }\href@noop {}
  {\bibfield  {journal} {\bibinfo  {journal} {Science}\ }\textbf {\bibinfo
  {volume} {343}},\ \bibinfo {pages} {1493--1496} (\bibinfo {year}
  {2014})}\BibitemShut {NoStop}%
\bibitem [{\citenamefont {Ashkin}\ and\ \citenamefont
  {Dziedzic}(1975)}]{Ashkin75}%
  \BibitemOpen
  \bibfield  {author} {\bibinfo {author} {\bibfnamefont {A.}~\bibnamefont
  {Ashkin}}\ and\ \bibinfo {author} {\bibfnamefont {J.~M.}\ \bibnamefont
  {Dziedzic}},\ }\bibfield  {title} {\enquote {\bibinfo {title} {Optical
  levitation of liquid drops by radiation pressure},}\ }\href@noop {}
  {\bibfield  {journal} {\bibinfo  {journal} {Science}\ }\textbf {\bibinfo
  {volume} {187}},\ \bibinfo {pages} {1073--1075} (\bibinfo {year}
  {1975})}\BibitemShut {NoStop}%
\bibitem [{\citenamefont {Berry}\ and\ \citenamefont {Geim}(1997)}]{Berry97}%
  \BibitemOpen
  \bibfield  {author} {\bibinfo {author} {\bibfnamefont {M.V.}\ \bibnamefont
  {Berry}}\ and\ \bibinfo {author} {\bibfnamefont {A.K.}\ \bibnamefont
  {Geim}},\ }\bibfield  {title} {\enquote {\bibinfo {title} {Of flying frogs
  and levitrons},}\ }\href@noop {} {\bibfield  {journal} {\bibinfo  {journal}
  {Euro. Phys. S.}\ }\textbf {\bibinfo {volume} {18}},\ \bibinfo {pages} {307}
  (\bibinfo {year} {1997})}\BibitemShut {NoStop}%
\bibitem [{\citenamefont {Ikezoe}\ \emph {et~al.}(1998)\citenamefont {Ikezoe},
  \citenamefont {Hirota}, \citenamefont {Nakagawa},\ and\ \citenamefont
  {Kitazawa}}]{Ikezoe98}%
  \BibitemOpen
  \bibfield  {author} {\bibinfo {author} {\bibfnamefont {Y.}~\bibnamefont
  {Ikezoe}}, \bibinfo {author} {\bibfnamefont {N}~\bibnamefont {Hirota}},
  \bibinfo {author} {\bibfnamefont {J.}~\bibnamefont {Nakagawa}}, \ and\
  \bibinfo {author} {\bibfnamefont {K}~\bibnamefont {Kitazawa}},\ }\bibfield
  {title} {\enquote {\bibinfo {title} {Making water levitate},}\ }\href@noop {}
  {\bibfield  {journal} {\bibinfo  {journal} {Nature}\ }\textbf {\bibinfo
  {volume} {393}},\ \bibinfo {pages} {749--750} (\bibinfo {year}
  {1998})}\BibitemShut {NoStop}%
\bibitem [{\citenamefont {Dhariwal}\ \emph {et~al.}(1993)\citenamefont
  {Dhariwal}, \citenamefont {Hall},\ and\ \citenamefont {Ray}}]{Dhariwal93}%
  \BibitemOpen
  \bibfield  {author} {\bibinfo {author} {\bibfnamefont {V.}~\bibnamefont
  {Dhariwal}}, \bibinfo {author} {\bibfnamefont {P.~G.}\ \bibnamefont {Hall}},
  \ and\ \bibinfo {author} {\bibfnamefont {A.~K.}\ \bibnamefont {Ray}},\
  }\bibfield  {title} {\enquote {\bibinfo {title} {Measurements of collection
  efficiency of single, charged droplets suspended in a stream of submicron
  particles with an electrodynamic balance},}\ }\href@noop {} {\bibfield
  {journal} {\bibinfo  {journal} {J. Aerosol Sci.}\ }\textbf {\bibinfo {volume}
  {24}},\ \bibinfo {pages} {197 -- 209} (\bibinfo {year} {1993})}\BibitemShut
  {NoStop}%
\bibitem [{\citenamefont {Kelling}\ and\ \citenamefont
  {Wurm}(2009)}]{Kelling09}%
  \BibitemOpen
  \bibfield  {author} {\bibinfo {author} {\bibfnamefont {T.}~\bibnamefont
  {Kelling}}\ and\ \bibinfo {author} {\bibfnamefont {G.}~\bibnamefont {Wurm}},\
  }\bibfield  {title} {\enquote {\bibinfo {title} {Self-sustained levitation of
  dust aggregate ensembles by temperature-gradient-induced overpressures},}\
  }\href@noop {} {\bibfield  {journal} {\bibinfo  {journal} {Phys. Rev. Lett.}\
  }\textbf {\bibinfo {volume} {103}},\ \bibinfo {pages} {215502} (\bibinfo
  {year} {2009})}\BibitemShut {NoStop}%
\bibitem [{\citenamefont {Knudsen}(1909)}]{Knudsen09}%
  \BibitemOpen
  \bibfield  {author} {\bibinfo {author} {\bibfnamefont {M.}~\bibnamefont
  {Knudsen}},\ }\bibfield  {title} {\enquote {\bibinfo {title} {Eine revision
  der gleichgewichtsbedingung der gase. {T}hermische molekularstr{\"o}mung.}}\
  }\href@noop {} {\bibfield  {journal} {\bibinfo  {journal} {Ann. Phys.}\
  }\textbf {\bibinfo {volume} {336}},\ \bibinfo {pages} {205} (\bibinfo {year}
  {1909})}\BibitemShut {NoStop}%
\bibitem [{\citenamefont {Kelling}\ \emph {et~al.}(2011)\citenamefont
  {Kelling}, \citenamefont {Wurm},\ and\ \citenamefont {Dürmann}}]{Kelling11}%
  \BibitemOpen
  \bibfield  {author} {\bibinfo {author} {\bibfnamefont {T.}~\bibnamefont
  {Kelling}}, \bibinfo {author} {\bibfnamefont {G.}~\bibnamefont {Wurm}}, \
  and\ \bibinfo {author} {\bibfnamefont {C.}~\bibnamefont {Dürmann}},\
  }\bibfield  {title} {\enquote {\bibinfo {title} {Ice particles trapped by
  temperature gradients at mbar pressure},}\ }\href@noop {} {\bibfield
  {journal} {\bibinfo  {journal} {Rev. Sci. Instrum.}\ }\textbf {\bibinfo
  {volume} {82}},\ \bibinfo {eid} {115105} (\bibinfo {year}
  {2011})}\BibitemShut {NoStop}%
\bibitem [{\citenamefont {Zheng}(2002)}]{Zheng02}%
  \BibitemOpen
  \bibfield  {author} {\bibinfo {author} {\bibfnamefont {F.}~\bibnamefont
  {Zheng}},\ }\bibfield  {title} {\enquote {\bibinfo {title} {Thermophoresis of
  spherical and non-spherical particles: a review of theories and
  experiments},}\ }\href@noop {} {\bibfield  {journal} {\bibinfo  {journal}
  {Adv. Colloid and Interface Sci.}\ }\textbf {\bibinfo {volume} {97}},\
  \bibinfo {pages} {255 -- 278} (\bibinfo {year} {2002})}\BibitemShut {NoStop}%
\bibitem [{\citenamefont {Kadoya}\ \emph {et~al.}(1985)\citenamefont {Kadoya},
  \citenamefont {Matsunaga},\ and\ \citenamefont {Nagashima}}]{Kadoya85}%
  \BibitemOpen
  \bibfield  {author} {\bibinfo {author} {\bibfnamefont {K.}~\bibnamefont
  {Kadoya}}, \bibinfo {author} {\bibfnamefont {N.}~\bibnamefont {Matsunaga}}, \
  and\ \bibinfo {author} {\bibfnamefont {A.}~\bibnamefont {Nagashima}},\
  }\bibfield  {title} {\enquote {\bibinfo {title} {Viscosity and thermal
  conductivity of dry air in the gaseous phase},}\ }\href@noop {} {\bibfield
  {journal} {\bibinfo  {journal} {J. Phys. Chem. Ref. Data}\ }\textbf {\bibinfo
  {volume} {14}},\ \bibinfo {pages} {947--970} (\bibinfo {year}
  {1985})}\BibitemShut {NoStop}%
\bibitem [{\citenamefont {Takata}\ \emph {et~al.}(1994)\citenamefont {Takata},
  \citenamefont {Aoki},\ and\ \citenamefont {Sone}}]{Takata94}%
  \BibitemOpen
  \bibfield  {author} {\bibinfo {author} {\bibfnamefont {S.}~\bibnamefont
  {Takata}}, \bibinfo {author} {\bibfnamefont {K.}~\bibnamefont {Aoki}}, \ and\
  \bibinfo {author} {\bibfnamefont {Y.}~\bibnamefont {Sone}},\ }\bibfield
  {title} {\enquote {\bibinfo {title} {Thermophoresis of a sphere with a
  uniform},}\ }\href@noop {} {\bibfield  {journal} {\bibinfo  {journal} {Prog.
  Astronaut. Aeronaut.}\ }\textbf {\bibinfo {volume} {159}},\ \bibinfo {pages}
  {626} (\bibinfo {year} {1994})}\BibitemShut {NoStop}%
\bibitem [{\citenamefont {Yamamoto}\ and\ \citenamefont
  {Ishihara}(1988)}]{Yamamoto88}%
  \BibitemOpen
  \bibfield  {author} {\bibinfo {author} {\bibfnamefont {K.}~\bibnamefont
  {Yamamoto}}\ and\ \bibinfo {author} {\bibfnamefont {Y.}~\bibnamefont
  {Ishihara}},\ }\bibfield  {title} {\enquote {\bibinfo {title} {Thermophoresis
  of a spherical particle in a rarefied gas of a transition regime},}\
  }\href@noop {} {\bibfield  {journal} {\bibinfo  {journal} {Phys. Fluids}\
  }\textbf {\bibinfo {volume} {31}},\ \bibinfo {pages} {3618--3624} (\bibinfo
  {year} {1988})}\BibitemShut {NoStop}%
\bibitem [{\citenamefont {Eckerskorn}\ \emph {et~al.}(2015)\citenamefont
  {Eckerskorn}, \citenamefont {Bowman}, \citenamefont {Kirian}, \citenamefont
  {Awel}, \citenamefont {Wiedorn}, \citenamefont {K\"upper}, \citenamefont
  {Padgett}, \citenamefont {Chapman},\ and\ \citenamefont
  {Rode}}]{Eckerskorn15}%
  \BibitemOpen
  \bibfield  {author} {\bibinfo {author} {\bibfnamefont {N.}~\bibnamefont
  {Eckerskorn}}, \bibinfo {author} {\bibfnamefont {R.}~\bibnamefont {Bowman}},
  \bibinfo {author} {\bibfnamefont {R.~A.}\ \bibnamefont {Kirian}}, \bibinfo
  {author} {\bibfnamefont {S.}~\bibnamefont {Awel}}, \bibinfo {author}
  {\bibfnamefont {M.}~\bibnamefont {Wiedorn}}, \bibinfo {author} {\bibfnamefont
  {J.}~\bibnamefont {K\"upper}}, \bibinfo {author} {\bibfnamefont {M.~J.}\
  \bibnamefont {Padgett}}, \bibinfo {author} {\bibfnamefont {H.~N.}\
  \bibnamefont {Chapman}}, \ and\ \bibinfo {author} {\bibfnamefont {A.~V.}\
  \bibnamefont {Rode}},\ }\bibfield  {title} {\enquote {\bibinfo {title}
  {Optically induced forces imposed in an optical funnel on a stream of
  particles in air or vacuum},}\ }\href@noop {} {\bibfield  {journal} {\bibinfo
   {journal} {Phys. Rev. Applied}\ }\textbf {\bibinfo {volume} {4}},\ \bibinfo
  {pages} {064001} (\bibinfo {year} {2015})}\BibitemShut {NoStop}%
\bibitem [{\citenamefont {Rohatschek}(1995)}]{Rohatschek95}%
  \BibitemOpen
  \bibfield  {author} {\bibinfo {author} {\bibfnamefont {H.}~\bibnamefont
  {Rohatschek}},\ }\bibfield  {title} {\enquote {\bibinfo {title}
  {Semi-empirical model of photophoretic forces for the entire range of
  pressures},}\ }\href@noop {} {\bibfield  {journal} {\bibinfo  {journal} {J.
  Aerosol Sci.}\ }\textbf {\bibinfo {volume} {26}},\ \bibinfo {pages} {717 --
  734} (\bibinfo {year} {1995})}\BibitemShut {NoStop}%
\bibitem [{\citenamefont {Golshtein}\ and\ \citenamefont
  {Elperin}(1996)}]{Golshtein1996}%
  \BibitemOpen
  \bibfield  {author} {\bibinfo {author} {\bibfnamefont {E.}~\bibnamefont
  {Golshtein}}\ and\ \bibinfo {author} {\bibfnamefont {T.}~\bibnamefont
  {Elperin}},\ }\bibfield  {title} {\enquote {\bibinfo {title} {Convective
  instabilities in rarefied gases by direct simulation monte carlo method},}\
  }\href@noop {} {\bibfield  {journal} {\bibinfo  {journal} {J. Thermophys Heat
  Transfer}\ }\textbf {\bibinfo {volume} {10}},\ \bibinfo {pages} {250--256}
  (\bibinfo {year} {1996})}\BibitemShut {NoStop}%
\bibitem [{\citenamefont {Stefanov}\ \emph {et~al.}(2002)\citenamefont
  {Stefanov}, \citenamefont {Roussinov},\ and\ \citenamefont
  {Cercignani}}]{Stefanov2002}%
  \BibitemOpen
  \bibfield  {author} {\bibinfo {author} {\bibfnamefont {S.}~\bibnamefont
  {Stefanov}}, \bibinfo {author} {\bibfnamefont {V.}~\bibnamefont {Roussinov}},
  \ and\ \bibinfo {author} {\bibfnamefont {C.}~\bibnamefont {Cercignani}},\
  }\bibfield  {title} {\enquote {\bibinfo {title} {Rayleigh-{B}\'{e}nard flow
  of a rarefied gas and its attractors. {I}. {C}onvection regime},}\
  }\href@noop {} {\bibfield  {journal} {\bibinfo  {journal} {Phys. Fluids}\
  }\textbf {\bibinfo {volume} {14}},\ \bibinfo {pages} {2255--2269} (\bibinfo
  {year} {2002})}\BibitemShut {NoStop}%
\end{thebibliography}%

\end{document}